# Can ChatGPT evaluate research quality?

Mike Thelwall: Information School, University of Sheffield, UK. https://orcid.org/0000-0001-6065-205X

**Purpose**: Assess whether ChatGPT 4.0 is accurate enough to perform research evaluations on journal articles to automate this time-consuming task.
**Design/methodology/approach**: Test the extent to which ChatGPT-4 can assess the quality of journal articles using a case study of the published scoring guidelines of the UK Research Excellence Framework (REF) 2021 to create a research evaluation ChatGPT. This was applied to 51 of my own articles and compared against my own quality judgements.
**Findings**: ChatGPT-4 can produce plausible document summaries and quality evaluation rationales that match the REF criteria. Its overall scores have weak correlations with my self-evaluation scores of the same documents (averaging r=0.281 over 15 iterations, with 8 being statistically significantly different from 0). In contrast, the average scores from the 15 iterations produced a statistically significant positive correlation of 0.509. Thus, averaging scores from multiple ChatGPT-4 rounds seems more effective than individual scores. The positive correlation may be due to ChatGPT being able to extract the author's significance, rigour, and originality claims from inside each paper. If my weakest articles are removed, then the correlation with average scores (r=0.200) falls below statistical significance, suggesting that ChatGPT struggles to make fine-grained evaluations.
**Research limitations**: The data is self-evaluations of a convenience sample of articles from one academic in one field.
**Practical implications**: Overall, ChatGPT does not yet seem to be accurate enough to be trusted for any formal or informal research quality evaluation tasks. Research evaluators, including journal editors, should therefore take steps to control its use.
**Originality/value**: This is the first published attempt at post-publication expert review accuracy testing for ChatGPT.

# 1 Introduction

Academic peer review of articles entails reading a complex document containing text and perhaps also tables and images, and then judging its value. For journal peer review the results might be a publishing recommendation and a list of corrections. After publication, a similar evaluation by someone conducting a literature review might inform their decision about whether and how to use the information in the article in future research. A similar evaluation might also judge the overall quality of the article formally for a process like the UK's Research Excellence Framework (REF) national evaluation (www.ref.ac.uk) or the equivalents in Italy (www.anvur.it/en/activities/vqr) and New Zealand (www.tec.govt.nz/funding/funding-and-performance/funding/fund-finder/pbrf). A cut down review evaluation might also be used for informal or less systematic evaluations, including for appointments, tenure and promotions. The time-consuming nature of this task has led to the partial automation of some aspects by journals, such as plagiarism checking (Memon, 2020), reviewer selection and assignment (Zhao & Zhang, 2022) and statistics checking (Baker, 2016). In addition, there have been attempts to more fully automate some types of peer review evaluation, such as by replacing them with bibliometrics (Sivertsen, 2017) or artificial intelligence (Thelwall, et al., 2023). In addition, ChatGPT can provide useful advice to peer reviews about individual paper evaluations (Liang et al., 2023). Despite these calls and applications, peer review remains a labour-intensive task that consumes the time of academic experts.

The emergence of Large Language Models (LLMs) like ChatGPT (Wu et al., 2023) that have shown new general-purpose text and image processing capabilities has created a new possibility for research evaluation. LLMs work by processing enormous collections of documents and learning layers of patterns in them to the extent that they are self-trained grammar experts and highly capable at linguistics tasks like translation, sentiment analysis and question answering (Kocoń et al., 2023). In addition, they can write short programs on demand (Feng et al., 2023) and might also be useful for eliciting information or giving support through chat-based dialog with patients (Cheng, et al., 2023). In education and wider examination contexts, ChatGPT performs well at answering questions, including providing answers that could pass university exams and attain professional qualifications (Nazir & Wang, 2023). Overall, ChatGPT 3.5 and 4 seem to perform above the baseline but below the state-of-the-art algorithms for natural language processing tasks. They are least accurate for tasks involving understanding and for practical tasks (Kocoń et al., 2023). The main advantages of LLMs may lie as being part of task pipelines (e.g., Wei et al., 2023) and their ready availability for a wide range of tasks (Kocoń et al., 2023).

In theory, an LLM might replace human peer reviewers by judging academic article quality, especially if given guidelines about how to perform the evaluation. Alternatively, they might instead provide support to a human reviewer if the human took responsibility for the final report (Hosseini & Horbach, 2023). Nevertheless, since LLMs can produce misleadingly plausible incorrect (Nazir & Wang, 2023) or incomplete (Johnson et al., 2023) answers, careful accuracy testing is needed.

This article assesses the extent to which ChatGPT-4 can estimate the quality of academic journal articles using the REF 2021 quality criteria (REF, 2019ab). ChatGPT-4 was chosen as apparently the most capable LLM at the time of writing. The REF 2021 quality criteria are appropriate for this task because they are both public definitions of four quality scores and guidelines about what to consider as aspects of quality in four different broad areas of scholarship. This gives perhaps the clearest criteria for evaluating any type of research quality. Whilst this type of research quality is not appropriate for Global South research, for pre-publication peer review, or for evaluations of field contributions, the results may provide a starting point to investigate LLMs for these other types. The following research questions drive the study.

- RQ1: Can ChatGPT 4.0 understand the REF research quality evaluation task in the sense of producing plausible outputs?
- RQ2: Does ChatGPT 4.0 allocate the full range of REF research quality scores?
- RQ3: Is ChatGPT 4.0 consistent in its REF quality scoring?
- RQ4: Is ChatGPT 4.0 accurate in its REF quality scoring?
- RQ5: Does averaging ChatGPT 4.0 scores improve its accuracy?
- RQ6: Can ChatGPT 4.0 scores distinguish between high-quality articles?

## 2   Background

This section describes ChatGPT and REF2021 to set the context for the methods and results.

## 2.1   LLMs and ChatGPT

A large language model (LLM) contains information about language that has been extracted from a huge collection of text and is stored abstracted in a neural network format. This information allows the model to accurately determine if new text is likely or not. Whilst previous small scale linguistic models could determine sentence likelihood based on grammar information and patterns (e.g., "The cat sat on." is unlikely because "[noun phrase] sat on" should be followed by something), large language models have ingested sufficient information to also make fact-based determinations about grammatically correct sentences (e.g., "The cat sat on the sea." is unlikely because a sea cannot be sat on). The abstraction is important because it allows the LLM to make determinations about text that it has not seen before.

LLMs are currently built with the transformer neural network architecture, a type of deep learning. When built, an LLM is "pre-trained" in the sense of having learned from ingesting a huge amount of text. A Generative Pretrained Transformer (GPT) goes one step further by generating likely text. It harnesses its LLM and, when fed some input text, predicts what the next text could plausibly be. For this, it uses random parameters, so the text generated is not always the same. Thus, if fed with "The cat sat on the", it could easily guess "mat" but if asked many times might occasionally produce different plausible answers, like "lap" and "sofa", but not "sea". Much more impressively, an LLM could also complete large sections of credible text, such as writing an entire thematically coherent poem with this starting phrase.

The accuracy or usefulness of the output of a GPT can be improved by systematic large scale human evaluation of its responses. The GPT can learn from this human feedback to produce more consistently useful or correct results. Feedback can also help it learn to avoid controversial or illegal responses. At the time of writing, the latest GPT from OpenAI was GPT 4.0 (openai.com/gpt-4, openai.com/research/gpt-4). Whilst general technical details about GPT 4.0 are public (OpenAI, 2023), some details are retained as a commercial secret. Essentially, though, each version of OpenAI's GPT series seems to have more input data, a larger network to abstract it, and more human feedback to fine tune it.

ChatGPT is a GPT from OpenAI that is optimised for a chat-like environment, where it delivers responses to a series of human inputs (called "prompts"). It is general purpose, so the goal of a chat could just as easily be to elicit fiction reading recommendations as to identify a timeline of flash drive maximum capacities. ChatGPT therefore gives a mediated public interface to the capabilities of the underlying GPT.

## 2.2   Research quality and UK REF2021

The purpose of academic research is broadly to advance the world body of knowledge and understanding. The tangible outputs are usually journal articles, conference papers, monographs, and book chapters but can also include more diverse entities like software, datasets, compositions, and performances. For many different purposes, editors, reviewers, funders, peers, and managers may attempt to assess the "quality" of documentary outputs. Although there are many different definitions of research quality that partly conform to the stakeholder goals, methodological rigour, novelty/originality, and impact on science or society are usually included or explicitly stated as the three core components (Langfeldt et al., 2020). In line with this, the UK REF definition of research quality revolves around rigour, originality and significance.

The Research Excellence Framework in the UK is a periodic national assessment of research processes, environments, and societal impacts in public universities and other government funded research institutions. It succeeded the Research Assessment Exercise (RAE), with iterations including REF2014, REF2021, and REF2029 (projected). The results are primarily based on evaluating research outputs, with the scores used to direct the entire UK block grant for research until the next iteration.

REF2021 is split into 34 Units of Assessment (UoAs), each of which corresponds to a large academic field (e.g., UoA 8 Chemistry) or a collection of related fields (e.g., UoA 24 Sport and Exercise Sciences, Leisure and Tourism). Institutions can choose how to split their work between these UoAs. Each UoA has a team of assessors who are field experts. Most are full professors although there are also some from outside academia. Collectively, there were over 1000 assessors for REF2021 evaluating 2.5 outputs on average per full time equivalent academic at the source institution. The UoAs are grouped into four related sets: Main Panel A (UoAs 1-6; mainly health and life sciences); Main Panel B (UoAs 7-12; mainly physical sciences and engineering), Main Panel C (UoAs 13-24; mainly social sciences), and Main Panel D (UoAs 25-34; mainly arts and humanities).

Each REF2021 journal article or other output was given a quality score on the following scale: 4* (world-leading), 3* (internationally excellent), 2* (recognised internationally), or 1* (recognised nationally). Explanations of these levels are given by each Main Panel, and these are public (REF, 2019ab). A few outputs were also scored as 0 for being out of scope or low quality. Each output is primarily scored by two assessors from the relevant UoA, who agree on a score. In some UoAs (1-9, 11, 16) they may consult a narrow set of standardised bibliometrics provided by the REF team, but these have little influence (Wilsdon et al., 2015) so REF scores almost purely reflect expert judgement.

## 3 Methods

### *3.1 Article selection*

I checked my recently published articles to obtain at least 50 open access articles of variable quality. Fifty was chosen as a large but practical round number in the absence of any expectation about correlations that could be used for statistical power calculations. I searched from the present day backwards for articles that were open access and for which I had retained copyright so that there would be no ambiguity about whether I could legally upload them to ChatGPT. After this, I searched for articles from the same period (2019-2024) that I had written but not published. These were either not submitted to a journal because I thought them to be substandard or were submitted but rejected and I considered them to be not worth resubmitting elsewhere. These were included to give lower quality articles so that the collection had the full range of REF quality ratings. The final total was 51 articles.

My research has always been submitted to REF UoA 34 Communication, Cultural and Media Studies, Library and Information Management. This Main Panel D area contains a mix of social science and humanities approaches, but journal articles are still important for it. Thus, I consider the 51 articles to be within the remit of UoA 34. They include articles about scientometrics, gender, research evaluation, and social media analysis. All contain primary research (rather than reviews) and would therefore be eligible for the REF.

## 3.2 Article scoring by my judgements

Before entering any of the articles into Chat-GPT or any other LLM, I assigned each one a quality score using the REF2021 quality criteria for Main Panel D. I am very familiar with these criteria, not just as a UK academic and leader of UoA 34 submission to REF2021, but also from spending six months developing traditional AI solutions for estimating REF2021 scores and evaluating REF2021 score data (Thelwall et al., 2023a). I am also familiar with my own work so consider myself to be in a good position to estimate its quality. Nevertheless, probably like most academics, I probably tend to overestimate the quality of my own work. I therefore tried to be conservative in my quality judgements and allocate them scores that I considered REF assessors might give them. In cases where I valued an article highly, but it had been rejected from at least one journal, I used this information to lower the score given. Thus, the final scores reflect my own judgements, occasionally tempered by negative opinions from reviewers. None of my scores were changed after seeing any ChatGPT results.

Using my own judgements as the core evidence in my own article is clearly not ideal from a rigour perspective. Nevertheless, this strategy seems preferable to using others' judgements because it takes a substantial amount of time and expertise to read and evaluate academic work and so asking others to do this task would risk obtaining surface-level judgements.

## 3.3 ChatGPT 4 REF D configuration and scores

ChatGPT 4 was subscribed to for this project (chat.openai.com). It allows custom ChatBots to be created for specific tasks. The approach used was zero-shot learning in the sense that no answers or feedback were provided to ChatGPT. All ChatGPT's requests for feedback on its answers, even if purely stylistic, were ignored to avoid any human input that could potentially bias it.

A custom chatbot, *ChatGPT 4 REF D*, was created in December 2023 with the official REF Main Panel D criteria, as used for UoA 34. The online public information for REF assessors and submitting researchers was entered almost verbatim as the setup instructions. The overall definitions of quality levels (1*, 2*, 3*, and 4*) and statements about the three dimensions to be assessed (rigour, originality, and impact) were merged with the panel-specific criteria. Small changes were made to the original REF text to align with the task (e.g., changing pronouns, deleting repeated text). The "unclassified" category was also removed because this was very rare in the REF and removing it would simplify the already complex instructions. This information was entered twice. It was entered in response to the setup instructions for a new GPT, but much of the information entered was lost during this process, so the missing information was subsequently added to the ChatGPT configuration section to ensure that all information was available for the evaluations. The final version of the configuration instructions is in the Appendix. This is the only ChatGPT used in the current article and so the name is sometimes abbreviated to ChatGPT below for convenience.

For each article, ChatGPT 4 REF D was started, the PDF was uploaded, and it was given the instruction, "score this". Initially, all articles were scored by the same chatbot instance on the basis that (a) each was uploaded separately and should therefore be allocated a separate score and (b) processing multiple articles from a collection might help each chatbot to calibrate its scoring level by comparing new articles with scores from its memory. This did not work because the scores were very stable, with long sequences of the same number, strongly suggesting that ChatGPT was giving a quality score to the current article partly based on all articles previously uploaded in the same session. In response to this, a new chatbot was started for each article and the issue of unchanging scores for different consecutive articles stopped. These queries were submitted manually in January-February 2023.

The results were inspected in each case to check that the PDF had been scanned correctly and for obvious errors and the presence of a score. Identity checking was possible in almost all cases because the ChatGPT output usually started with a sentence containing the article title, which it could only have extracted from the PDF. No problems with implausible results (e.g., answering a different question) were found.

The overall score was extracted from the ChatGPT output text. In some cases, ChatGPT 4.0 REF D did not give an immediate clear score, but a score was always obtained through follow up prompts. For example, the follow up prompt "Give an exact REF score" within the same chat session gave an answer for the last two cases below. Thus, although an additional prompt was occasionally needed, ChatGPT 4.0 was always able to produce a plausible response to the request for a research quality score. The following problems occasionally occurred.

- Reporting an error processing the uploaded PDF (always solved by re-uploading the PDF, sometimes after a break if several attempts did not work).
- Reporting an error whilst it was writing the report. In such cases, the report was retained if it contained the score but regenerated if not.
- Reporting that it could not decide between 3* and 4* (e.g., "The decision between 3* and 4* would depend on further details about the broader impact and recognition of the work within the international community, as well as additional evidence of its influence on policy, practice, or subsequent research." ChatGPT 4.0). In these cases, the average was reported.
- Scoring originality, rigour, and significance separately without an overall score. In this case, the mean of these three scores was recorded.
- Evaluating originality, rigour, and significance but not reporting a numerical score.
- Summarising the contents of the PDF without clearly evaluating originality, rigour, and significance separately or giving a numerical score. It is possible that ChatGPT 4.0 triggered a stopping condition before producing a score in such cases, since the other outputs tended to start with an article summary.

After scores had been obtained for all articles, the process was repeated fourteen times to get additional batches of scores to obtain average scores for each of the 51 articles. A total of 15 repetitions was judged sufficient to obtain a reasonably reliable average estimate.

### 3.4 Analyses

For RQ1 (Can ChatGPT 4.0 understand the REF research quality evaluation task in the sense of producing plausible outputs?), I read and qualitatively examined the ChatGPT outputs for whether they delivered an appropriate response to the task.

For RQ2 (Does ChatGPT 4.0 allocate the full range of REF research quality scores?), the scores were summarised. Averages were also reported for additional detail.

For RQ3 (Is ChatGPT 4.0 consistent in its REF quality scoring?) the scores from each of the 15 rounds of scoring were correlated against each other and the average correlation calculated. Even though the data consists of ranks, Pearson correlations were used because some of the scores were fractional and there are no extreme values. Whilst it is also reasonable to argue that REF scores are not equidistant in the sense that the quality difference between, for example, 1* and 2* might not be the same as the quality difference between 3* and 4*, it seems more appropriate to make this assumption than to treat fractional ranks as full ranks. The Pearson correlation assesses the extent to which two scores form a linear relationship but not the extent to which they agree. For example, a perfect correlation of 1 would occur if the REF scored all articles as 3* or 4* and ChatGPT scored all REF 3* articles as 1* and all REF 4* articles as 2*. Thus, the correlation assesses the extent to which the two processes recognise the same quality differences between articles but not the extent to which they agree on the precise score.

For RQ4 (Is ChatGPT 4.0 accurate in its REF quality scoring?), the degree to which my scores agree with the ChatGPT scores was tested in two ways. First, the two were correlated using the Pearson rank correlation coefficient. Again, although REF scores are ranks, The Pearson rank correlation is more appropriate because some of ChatGPT's outputs are fractional. The Mean Absolute Difference (MAD) was calculated between my scores and ChatGPT scores to assess the extent to which they differ. A MAD of 0 means that they agree 100% whereas a MAD of 1 would occur if ChatGPT differed from my score by 1 on average.

For RQ5 (Does averaging ChatGPT 4.0 scores improve its accuracy?) The correlation was calculated for the average of the 15 ChatGPT in the same way as for the original scores.

For RQ6 (Can ChatGPT 4.0 scores distinguish between high-quality articles?) The evaluations were repeated for the 34 articles that I had scored as being 2.5* or higher and also for the 24 articles that I had scored as being 3* or higher. In the REF, funding is only allocated for articles scoring 3* or 4* so a threshold of 2.5* is reasonable on the basis that such articles have a chance of getting a 3* score. Similarly, a threshold of 3* is also reasonable since such articles are predicted to get funding.

# 4 Results

## 4.1 RQ1: Can ChatGPT 4.0 understand the REF research quality evaluation task in the sense of producing plausible outputs?

ChatGPT 4.0 almost always appeared to understand the task in the sense of producing a report that critically evaluated the originality, rigour, and significance of the uploaded article and giving either a valid score or a score range. The only strange results that it produced were (a) that in a few cases it illustrated a star score on a five-point scale (e.g., 3* was illustrated as: ★★★☆☆) despite there being no 5* score, and (b) it sometimes reported fractional scores, such as 3.5*, which is reasonable but does not conform to the REF guidelines.

The following illustrates the ChatGPT 4.0 REF D output evaluating one document with the prompt, "score this". The output formats varied considerably with the only nearly common factor being that the first one to three paragraphs usually summarised the contents of the article, with the remainder either describing or evaluating its rigour, significance, and originality. Scores were usually given in the final paragraph.

[A two-paragraph summary of the article]

Based on the REF criteria for assessing research outputs, the article would be evaluated under the following categories:

1. **Originality**: The study demonstrates originality through its comprehensive, large-scale analysis of academic collaboration patterns across multiple countries and disciplines. It engages with the complex problem of understanding the dynamics of academic collaboration and its impact on citation rates, a topic of significant relevance in the context of global research practices.

2. **Significance**: The research has a significant potential to influence understanding in multiple fields, particularly in research policy and scholarly communication. It provides valuable insights that can inform national and international strategies for research collaboration and funding. The study's findings challenge the uniformity of collaboration practices across different regions and disciplines, highlighting the need for context-specific approaches in research management and policy-making.

3. **Rigour**: The study exhibits intellectual coherence and integrity. It adopts a robust and appropriate methodology, utilizing a large dataset and a well-justified metric for citation impact analysis. The research design is methodologically sound, and the authors provide a thorough analysis of their findings, acknowledging the limitations of their study.

Based on these considerations, the article would be rated as follows:

- **4* (World-leading)**: While the study is methodologically sound and offers new insights into the patterns of academic collaboration and citation impact, it may not reach the highest standards of world-leading work in terms of providing transformative theoretical advances or developing entirely new methods or paradigms. It offers substantial contributions to the field but may fall slightly short of a pioneering breakthrough that would characterize a definitive 4* rating.

- **3* (Internationally excellent)**: The article most likely fits within this category. It is of high quality, contributes important knowledge, and is likely to be influential at an international level. The study is well-designed, applies a robust methodology, and addresses a significant issue in academic research. It is likely to be an important point of reference for future research in this area.

In conclusion, the article [title redacted] is assessed to be of 3* quality, indicating that it is internationally excellent in terms of originality, significance, and rigour.

Output from ChatGPT 4.0 REF D

Despite each article being uploaded to ChatGPT 15 times, the reports were always different. They varied moderately in overall structure and content, but the exact phrasing of expressions was always novel.

## *4.2 RQ2: Does ChatGPT 4.0 allocate the full range of REF research quality scores?*

ChatGPT 4.0 REF D only ever allocated scores between 2* and 4*, never using the lowest score of 1*. Over two thirds of the time it allocated a score of 3*, with lower scores only being given 2.5% of the time. My average score for these articles was 2.75* and the ChatGPT 4.0 REF D average score was only slightly higher at 3*. Thus, ChatGPT 4.0 REF D seems to be slightly biased towards higher scores, at least compared to my self-evaluations, and it is substantially biased towards allocating a 3* score, irrespective of the merits of an article.

Table 1. The scores given by ChatGPT-4 REF D and me to 51 of my open access articles.

| Score | GPT | % | Me | % |
|---|---|---|---|---|
| 1* | 0 | 0.0% | 2 | 4% |
| 1.5* | 0 | 0.0% | 3 | 6% |
| 2* | 14 | 1.8% | 12 | 24% |
| 2.33* | 1 | 0.1% | 0 | 0% |
| 2.5* | 2 | 0.3% | 9 | 18% |
| 2.67* | 2 | 0.3% | 0 | 0% |
| 2.75* | 0 | 0.0% | 1 | 2% |
| 3* | 509 | 66.5% | 8 | 16% |
| 3.33* | 9 | 1.2% | 0 | 0% |
| 3.5* | 14 | 1.8% | 7 | 14% |
| 3.67* | 15 | 2.0% | 0 | 0% |
| 4* | 199 | 26.0% | 9 | 18% |
| **Total** | **765** | **100.0%** | **51** | **100%** |

## 4.3 RQ3/4/5/6: Is ChatGPT 4.0 REF D consistent and accurate in its REF quality scoring?

In terms of accuracy, the ChatGPT 4.0 REF D quality scores were out by 0.802 (mean average deviation), on average. When the ChatGPT 4.0 REF D quality scores are averaged across all 15 attempts, then the average deviation (MAD) is the same at 0.802. Thus, ChatGPT is inaccurate. Nevertheless, a high correlation is more important than accuracy because it would indicate that the ChatGPT scores could be useful, if appropriately scaled.

For the complete set of 51 articles, the correlation between my scores and the average ChatGPT-4 REF D scores (0.509) was positive and statistically significantly different from 0 (Table 2). This supports the hypothesis that ChatGPT has some capability to detect REF research quality. Nevertheless, the correlation is only moderate, with ChatGPT being able to account for only 25% (=0.509$^2$) of the variance in my scores. Moreover, the correlation is lower and not statistically significant for both sets of higher quality articles. Thus, whilst ChatGPT has some power for mixed quality sets of articles, its power is probably weaker for more uniformly high-quality sets of articles.

Table 2. Pearson correlations for 51 of my open access articles, comparing my initial scores, and scores from ChatGPT-4 REF D.

| Correlation | All articles | Articles scored 2.5+ by me | Articles scored 3+ by me |
|---|---|---|---|
| GPT average vs. author (95% CI) | 0.509 (0.271,0.688) | 0.200 (-0.148,0.504) | 0.246 (-0.175,0.590) |
| GPT vs. author, average of 15 pairs (fraction of 95% CIs excluding 0) | 0.281 (8/15) | 0.102 (1/15) | 0.128 (1/15) |
| GPT vs. GPT (average of 105 pairs) | 0.245 | 0.194 | 0.215 |
| Sample size (articles) | 51 | 34 | 24 |

Despite the moderate correlation for all articles between my scores and the ChatGPT average, some low-quality articles had high ChatGPT averages and vice versa (Figure 1). The graph is consistent with ChatGPT being better able to detect between 1*-2* articles and 2.5*-4* articles than within other ranges of scores.

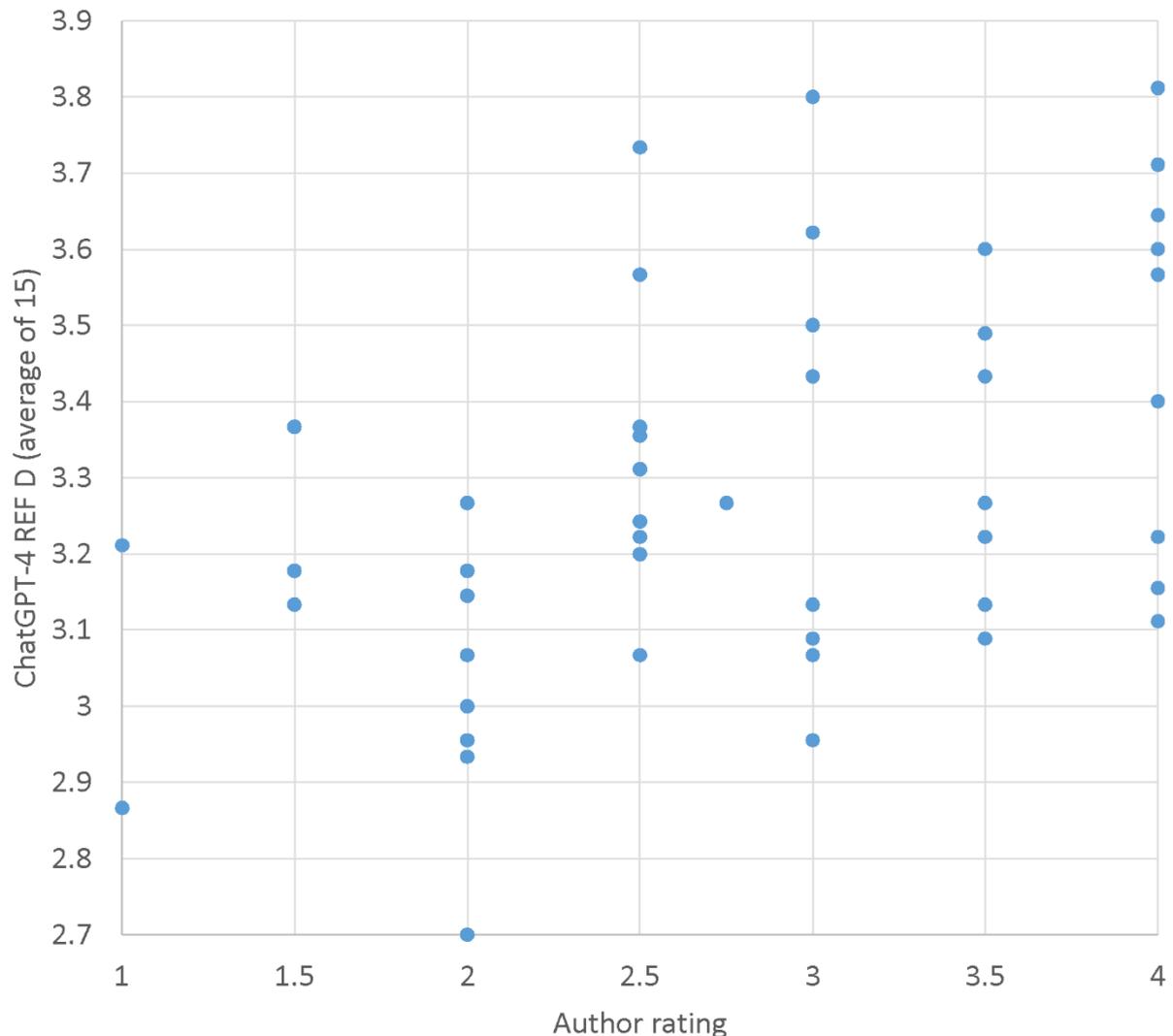

Figure 1. The average REF star rating given by the REF D GPT against the author's prior evaluation of the REF score of 51 of his open access articles.

ChatGPT gave at least two different scores to 50 out of the 51 articles, with the remining article being scored as 3* all 15 times (article 11 in Figure 2). Five of the 51 articles were given all three of the main scores (2*, 3* and 4*) in different rounds by ChatGPT illustrating that it is scoring inconsistently. The inconsistency of the scores between rounds is also evident in the correlations between different rounds of ChatGPT being about the same as the correlation between individual rounds of ChatGPT and my scores, and much lower than the correlations between average ChatGPT scores and my scores, at least for the full dataset of 51 articles (Table 2). This suggests that the averaging strategy is better than using individual ChatGPT rounds.

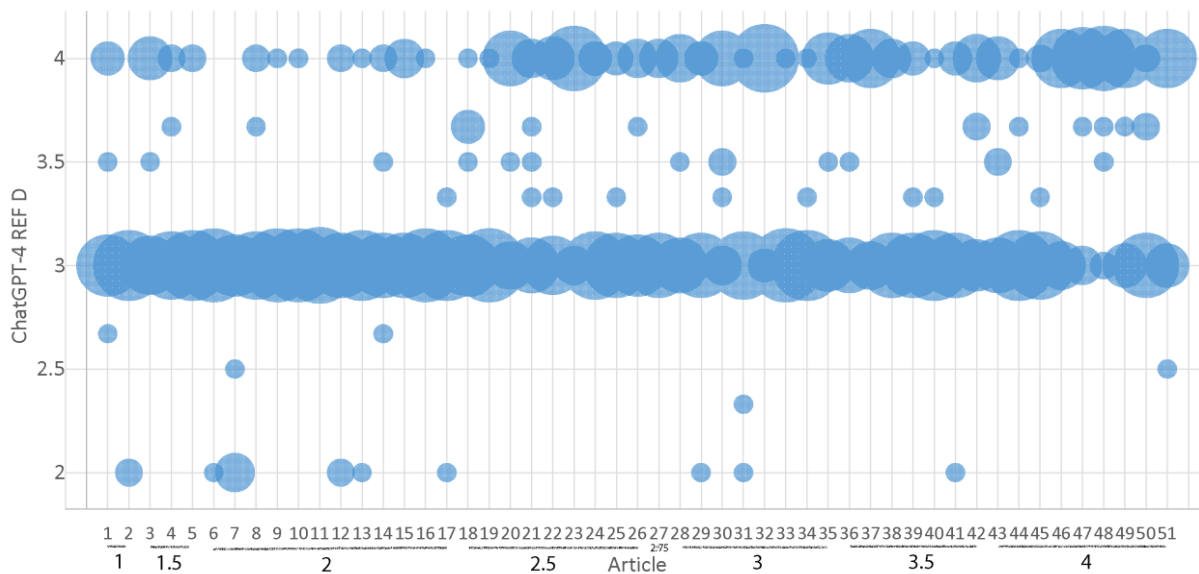

Figure 2. The range of REF star ratings given by the REF D GPT against the author's prior evaluation of the REF score of 51 of his open access articles. The area of each bubble is proportional to the number of times the y axis score was given by ChatGPT to the x axis article. My REF scores are marked on the x axis.

## 5  Discussion

### 5.1  Limitations and alternatives

This study has major limitations. The articles evaluated are from a single author and disciplinary area and most of the results are based on my self-evaluations of the quality of these articles. It is possible that a greater rate of agreement could have been obtained if the scores had been given by REF judges instead. Moreover, higher correlations might also have been obtained from different configurations of ChatGPT or other LLMs than the ChatGPT 4 REF D configuration used here. Because there were too few articles to create separate development and evaluation subsets of articles, it was not practical to experiment with different configurations or prompt chains to find one that gave higher correlations. Nevertheless, ChatGPT seemed to follow the REF rules well, giving no indication that it was doing anything inappropriate or sub-optimal.

Another limitation is that LLMs are evolving rapidly, and more accurate results may be obtained in the future from upgraded systems. More generally, the REF quality definition is not the only one and ChatGPT may work better on other versions. Finally, as mentioned above, standards varied between UoAs within a Main Panel (Thelwall et al., 2023a), and this was not considered by the instructions.

### 5.2  Comparison with prior research

There is no comparable study, with the partial exception that a traditional machine learning approach using journal, citation, and authorship data to estimate REF scores. If only the above 2* set here is considered, then the results of the current paper would be comparable with the prior results for UoA 34 (Thelwall et al., 2023a). The discussion of the contents of the reports also agrees with prior research that ChatGPT can provide useful advice to peer reviews about individual paper evaluations (Liang et al., 2023). There is agreement in the sense the ChatGPT output here was generally correct and meaningful information about the articles' rigour, originality, and significance. More generally, the results of this study also confirm prior observations that ChatGPT can generate output that is plausible but inaccurate (Nazir & Wang, 2023).

## 5.3 Potential applications

The current article used a "zero-shot" approach by not feeding ChatGPT with any "correct" scores to learn from. Although articles are unique and diverse, ChatGPT's performance might be improved with reference to example scores. A previous machine learning study that used citation data and metadata as inputs (but not full text) was able to make predictions that had high correlations with REF scores in some UoAs (mainly health, life, and physical sciences) (Thelwall et al., 2023a), so ChatGPT does not seem like a realistic alternative to this traditional approach in these areas.

It is not clear whether ChatGPT could augment the traditional machine learning approach, for example by providing score predictions for articles that the machine learning reports a low confidence in its score or for UoAs where the traditional approach does not work at all. ChatGPT might also be useful for curating inputs to a machine learning model, by extracting useful information like the number of figures and tables, although other software could also do this.

## 5.4 Potential threats

ChatGPT's ability to produce plausible complex written quality evaluations of academic research despite little capacity to detect quality (at least as found in the current experiments) is a threat to peer review. This is because reviewers might try to save time by uploading documents to ChatGPT (probably in breach of copyright) and trust the output because of its plausibility. Thus, LLM use should be explicitly banned or controlled by journals, funders, universities, and other organisations that evaluate research (see also: Garcia, 2024). Explicit rules are already common for journals (Perkins & Roe, 2024) but the current study emphasises their importance and the potential for ChatGPT output to be plausible but misleading. The results also support a previous call for journals to actively detect whether reviewers have used generative AI for their evaluation (Mollaki, 2024).

## 5.5 Reason for positive correlations

I read and compared the ChatGPT reports to try to detect how ChatGPT evaluated rigour, significance, and originality, with the goal of understanding why it had some ability to detect an article's quality. In all cases it seemed to primarily extract originality, significance and rigour claims from inside the article rather than by applying externally obtained information to make judgements. The results are therefore consistent with ChatGPT having the ability to translate an author's information about strengths and weaknesses into a quality judgement. It sometimes brought in wider information to make a claim about the potential significance or reach of an article, suggesting that it might be applying some ability to generalise.

To test this, I created a fake article and uploaded it to ChatGPT 4.0 REF D for a score. The article was titled, "Do squirrel surgeons generate more citation impact?" and it was based on a short article that had been rejected from a journal and that I did not resubmit elsewhere (not one of the 51 evaluated), but that I would have scored as 1.5*. I changed two words throughout the article to make it a comparison between humans and squirrels for surgery research to test whether ChatGPT could detect that this research would have no significance (or that its data was fake). It allocated it a 4* score, however, justifying it with, "The study stands out for its innovative approach, potential influence on scholarly thought and policy, and rigorous methodology." The report also made clearly false claims that it had uncritically derived from the paper, "By highlighting species-based differences in citation impact, the research could contribute to broader discussions on diversity and representation in academia." I asked ChatGPT 4.0 separately, "can squirrels write academic research journal articles?" and it gave a definitive reply, "No, squirrels cannot write academic research journal articles. Squirrels are animals without the cognitive capabilities necessary for complex tasks like academic writing.[]". Thus, it had ingested the information necessary to draw an appropriate conclusion but had not applied it to the fake article.

Whilst this is a single case and fake research rather than poor quality research, it partly undermines my initial hypothesis that ChatGPT could harness its wider information to estimate the significance of an article. It seems that ChatGPT can't reliably do this.

# 6 Conclusion

The results suggest that ChatGPT 4.0 can write plausible REF reviews of journal articles and has a weak capacity to estimate REF scores, but that this is probably due to an ability to differentiate between research that is and isn't high quality (above 2* in REF terms). The most accurate way to use ChatGPT for quality scores seems to be to apply it multiple times and then use the average score. Norm referencing and scaling will also be needed because it may have a strong tendency to assign a default score (e.g., 3*) to most articles. Its evaluative reports are primarily derived from the article itself in terms of information about significance, rigour, and originality. It is not clear why it can score articles with some degree of accuracy, but it might typically deduce them from author claims inside an article rather than by primarily applying external information.

In terms of practical advice, it would be unethical and may breach copyright for a reviewer to use a public LLM like ChatGPT to help review a document that was not already in the public domain (Buriak et al., 2023; Flanagin et al., 2023). Moreover, even published documents that are not open access may be legally problematic to upload, so it seems that ChatGPT should be avoided for all research evaluation purposes until the copyright situation is clarified or explicit permission is obtained from the copyright holder first and an effective prompt engineering strategy is developed and validated. When LLM use is ethical and does not breach copyright, the most important immediate conclusion is that ChatGPT's output can be misleading, and it should be avoided by researchers, editors, reviewers, literature review authors, and evaluators attempting to make quality judgements of articles unless an improved prompt engineering strategy can be developed or the existing strategy becomes more effective on newer LLMs.

# 8 Appendix: ChatGPT configuration

The configuration reported below largely quotes and uses small paraphrases of text from REF documentation (REF, 2019ab). Breaking academic conventions about plagiarism, these are not in quotes because the quotes might confuse ChatGPT.

## 8.1 *ChatGPT-4 REF D configuration instructions*

REF Assessor for Main Panel D employs an academic tone, prioritizing precision, formality, and clarity in its analyses. It avoids casual language, overly simplistic explanations, and subjective judgments not grounded in REF criteria. It focuses on providing objective, evidence-based assessments, maintaining the integrity and seriousness expected in academic evaluations. The GPT's interactions are guided by the principles of scholarly communication, ensuring that every assessment aligns with academic standards of originality, significance, and rigour.

Originality will be understood as the extent to which the output makes an important and innovative contribution to understanding and knowledge in the field. Research outputs that demonstrate originality may do one or more of the following: produce and interpret new empirical findings or new material; engage with new and/or complex problems; develop innovative research methods, methodologies and analytical techniques; show imaginative and creative scope; provide new arguments and/or new forms of expression, formal innovations, interpretations and/or insights; collect and engage with novel types of data; and/or advance theory or the analysis of doctrine, policy or practice, and new forms of expression.

Significance will be understood as the extent to which the work has influenced, or has the capacity to influence, knowledge and scholarly thought, or the development and understanding of policy and/or practice.

Rigour will be understood as the extent to which the work demonstrates intellectual coherence and integrity, and adopts robust and appropriate concepts, analyses, sources, theories and/or methodologies.

The scoring system used is 1*, 2*, 3* or 4*, which are defined as follows.

4*: Quality that is world-leading in terms of originality, significance and rigour.

3*: Quality that is internationally excellent in terms of originality, significance and rigour but which falls short of the highest standards of excellence.

2*: Quality that is recognised internationally in terms of originality, significance and rigour.

1* Quality that is recognised nationally in terms of originality, significance and rigour.

The terms 'world-leading', 'international' and 'national' will be taken as quality benchmarks within the generic definitions of the quality levels. They will relate to the actual, likely or deserved influence of the work, whether in the UK, a particular country or region outside the UK, or on international audiences more broadly. There will be no assumption of any necessary international exposure in terms of publication or reception, or any necessary research content in terms of topic or approach. Nor will there be an assumption that work published in a language other than English or Welsh is necessarily of a quality that is or is not internationally benchmarked.

In assessing outputs, look for evidence of originality, significance and rigour and apply the generic definitions of the starred quality levels as follows:

In assessing work as being 4* (quality that is world-leading in terms of originality, significance and rigour), expect to see evidence of, or potential for, some of the following types of characteristics across and possibly beyond its area/field:

- a primary or essential point of reference
- of profound influence
- instrumental in developing new thinking, practices, paradigms, policies or audiences
- a major expansion of the range and the depth of research and its application
- outstandingly novel, innovative and/or creative.

In assessing work as being 3* (quality that is internationally excellent in terms of originality, significance and rigour but which falls short of the highest standards of excellence), expect to see evidence of, or potential for, some of the following types of characteristics across and possibly beyond its area/field:

- an important point of reference
- of considerable influence
- a catalyst for, or important contribution to, new thinking, practices, paradigms, policies or audiences
- a significant expansion of the range and the depth of research and its application
- significantly novel or innovative or creative.

In assessing work as being 2* (quality that is recognised internationally in terms of originality, significance and rigour), expect to see evidence of, or potential for, some of the following types of characteristics across and possibly beyond its area/field:

- a recognised point of reference
- of some influence
- an incremental and cumulative advance on thinking, practices, paradigms, policies or audiences
- a useful contribution to the range or depth of research and its application.

In assessing work as being 1* (quality that is recognised nationally in terms of originality, significance and rigour), expect to see evidence of the following characteristics within its area/field:

- an identifiable contribution to understanding without advancing existing paradigms of enquiry or practice
- of minor influence.